\documentclass[useAMS,usenatbib]{mn2e}

\usepackage{graphicx}

\def\MV{\mbox{$M_{F606W}$}}

\def\mV{\mbox{$m_{F606W}$}}
\def\mI{\mbox{$m_{F814W}$}}
\def\mM{\mbox{$(m-M)_0$}}
\def\ebv{\mbox{$E(B-V)$}}
\def\feh{\mbox{$[Fe/H]$}}
\def\lage{\mbox{$log(\tau(yrs))$}}
\def\rc{\mbox{$r_{\rm core}$}}

\def\rh{\mbox{$r_{\rm h}$}}

\def\ds{\mbox{$d_\odot$}}

\def\cl{NGC\,6642 }

\def\hbi{\mbox{$HB_{index}$}}

\title[ACS photometry of the globular cluster NGC 6642]{The globular
cluster NGC 6642: \\ Evidence for a depleted mass function in a very old cluster}

\author[E. Balbinot et al.]{E. Balbinot${^1}$, B.X. Santiago$^{1}$, E. Bica$^{1}$ and C. Bonatto$^{1}$\\
$^1$Departamento de Astronomia, Universidade Federal do Rio Grande do Sul, Av. Bento
Gon\c{c}alves 9500\\ Porto Alegre 91501-970, RS, Brazil}

\begin{document}

\pagerange{\pageref{firstpage}--\pageref{lastpage}} 

\maketitle

\label{firstpage}

\begin{abstract}

We present photometry for the globular cluster \cl
using the F606W and F814W filters with the ACS/WFC third generation camera on 
board of Hubble Space Telescope. The Colour Magnitude
Diagram shows sources reaching $\approx 6~mags$ below the
turn-off in $\mV$. A theoretical isochrone fitting was performed and
evolutionary parameters were obtained, such as the metallicity $\feh = -1.80 
\pm 0.2$ and age $\lage = 10.14 \pm 0.05$. We confirm that \cl is located
in the Galactic bulge, with a distance to the Sun $\ds = 8.05 \pm 0.66 ~ kpc$ and
the reddening $\ebv = 0.46 \pm 0.02$. These values are in general agreement
with those of previous authors. About 30 blue stragglers were found within
the central 1.6 pc of \cl. They are strongly concentrated to the very central
regions. The cluster displays a well-developed horizontal branch,
with a much redder morphology than that of typical old halo globular 
clusters of similar metallicity. Completeness corrected
luminosity and mass functions were obtained for different annuli centred on \cl. 
Their spatial variation indicates the existence of
mass segregation and depletion of low mass stars. Most striking is the 
inverted shape of the mass function itself, with an increase in number 
as a function of increasing mass. This has been previously observed in other
globular clusters and is also the result of N-body simulations of 
stellar systems which have undergone $\simeq 90\%$ of their lifetime and
which are subjected to strong tidal effects.
We also analysed the density profile and concluded that \cl has a collapsed
core, provided completeness effects are correctly accounted for.
We thus conclude from independent means that \cl is a very old, 
highly-evolved, core-collapsed globular cluster with an atypical 
HB morphology. 
Its current location close to perigalactic, 
at only $1.4$ kpc from the Galactic
centre, may contribute to this high level of dynamical evolution and
stellar depletion.
\end{abstract}


\section{Introduction}

Given their relatively high contrast with the background,
the census of globular clusters (GCs) in the Galaxy is near completion.
The dense and highly extincted Galactic bulge, specially its 
inner regions, is the site where most missing clusters,
perhaps 10\% of the total known population,
are expected to be (Ivanov et al. 2005, 
Ortolani et al. 2006 and references therein). The bulge is known to host
old but fairly metal-rich stellar populations, with high $[\alpha/Fe]$
ratios, indicating that it formed on a short time-scale 
early in the Galactic history (Origlia et al. 2005). 
The bulge is thus an appropriate
region to look for ancient GCs with distinct metallicities
from those in the outer stellar halo.
The bulge environment may also cause extreme dynamical evolution of
star clusters due to bulge shocking and strong tidal 
effects \citep{aguilar,shin}.
For old enough GCs, effects of close stellar encounters,
which lead to mass segregation and stellar evaporation, are certain 
to be found in them as well \citep{baumgardt}.

Analyses of the structure and dynamical evolution of bulge GCs
require high-resolution imaging in order to resolve individual stars in
their cores and to effectively subtract contaminating field stars.
The Advanced Camera for Surveys in the Wide Field Channel (ACS/WFC) on board of
the Hubble Space Telescope (HST) has provided a significant leap in the
amount of such data on GCs, not only in the bulge, but also elsewhere
\citep{sar,richer,nath}. Recently, Marin-Franch et al. (2008) provided 
ACS photometry for 64 halo GCs and derived relative ages.
The ACS allows to study extremely dense fields using point 
source photometry. With such
technique it is feasible to use stellar interior models to
obtain evolutionary parameters of the cluster stellar population.

\cl (also designated by ESO 522-SC32 and GCl-97) 
is located at $l = 9.81^{\circ}$, $b = -6.44^{\circ}$, therefore
projected towards the Galactic bulge. It is thus superimposed onto bulge 
and inner halo stellar populations. Dynamical studies of this GC,
as well as other bulge clusters, have thus been extremely difficulty, 
due to limitations of ground based images.
Structural parameters were found by \citet{trager}, such as core radius 
$\rc = 6.1 \arcsec$, concentration parameter $c = 1.99$ and half light 
radius $\rh = 44 \arcsec$. They also conclude that \cl is a 
core-collapsed candidate.
Using spectroscopy of individual stars, \citet{minni} derived
$\feh = -1.40$. Later \citet{barbuy} using BVI photometry with SOAR, 
found $\feh = -1.3$, reddening $\ebv = 0.42 \pm 0.03$ and $\ds = 7.2
Kpc$ ($\mM=14.3$). In the catalogue compiled by \citet{harris},
similar results are quoted: $(m-M)_{V} = 15.90$, $\ebv = 0.41$ 
and $\feh = -1.35$. All these results are from ground-based studies. 
The only space-based dataset on \cl so far is that from 
the compilation of Colour-Magnitude Diagrams 
(CMDs) from \citet{pio}, who used the Wide Field Planetary
Camera (WFPC2) snapshots. 

In this work we present F606W (Broad V) and F814W (Broad I)
ACS photometry of \cl. Notice that \cl is not included in the sample
by Marin-Franch et al. (2008) since it is projected
towards the bulge.
We use this photometry not only to derive metallicity, reddening, 
distance and age, but also to make a detailed analysis of its e.g. structure,
including radial density profile (RDP), luminosity and mass functions,
horizontal branch (HB) morphology and blue stragglers. 
This paper is organised in the following way:
in Sect. 2 we describe the data and the reduction process. In
Sect. 3 we present the data analysis. In Sect. 4 there is a
brief discussion on the cluster age and evolution; we also present the
concluding remarks.

\section{Data}

\begin{figure}
   \includegraphics[width=80mm]{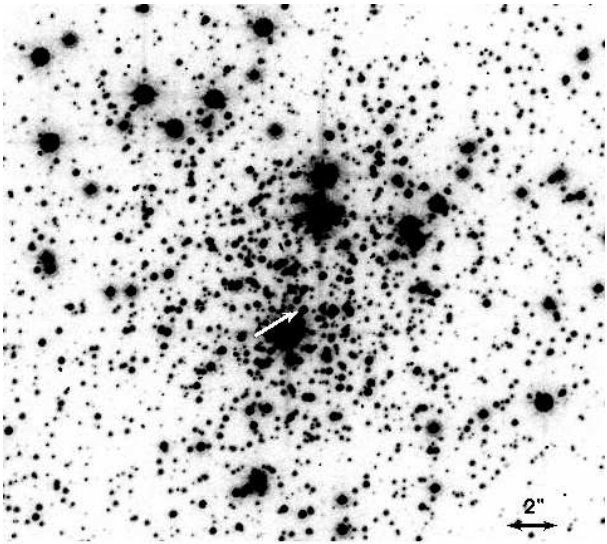}
   \caption{NGC6642: Central region (22 \arcsec$\times$21 \arcsec) of 
the F814W short exposure. North is up and East is right. The adopted centre 
is indicated by the white arrow. Notice the dense object, apparently formed by 
a clump of crowded stars just below the centre of this image. Even with 
ACS/HST the stars in the centre of \cl are not easily resolvable. }
   \label{cluster}
\end{figure}

The images were retrieved from the Space Telescope Science Institute
\footnote{http://www.stsci.edu/} (STScI) data archive and were automatically 
reduced by the STScI pipeline, i.e. they were corrected for 
bias, dark current and were divided by the flatfield image. The images 
of \cl are  part of the proposal 9799 and were obtained in 2004.
They cover a $202 \arcsec \times 202 \arcsec$ field of view at a spatial 
scale of $0.049\arcsec /pixel$, with a gap between the CCDs. 
Two exposures were taken in the F606W and F814W filters:
short ($10s$) and long ($340s$).
In Fig. \ref{cluster} we show the F814W short exposure for the cluster's 
central region. Notice the detached bright central object 
composed by unresolved individual stars located near the cluster centre.

The photometry was carried out using the {\tt DOLPHOT} software 
\citep{dolphot}. The following steps were taken to prepare the images for 
photometry: (i) the task {\tt ACSMASK} masks all bad pixels, (ii) the task 
{\tt SPLITGROUPS} splits the multi-extension {\tt FITS} image into single chip 
images. As described in the {\tt DOLPHOT} manual, a position reference image
must be adopted, preferentially a deep and uncrowded one. We used the F814W long 
exposure with geometric corrections applied (drizzled image) as the reference 
frame. The photometric measurements were performed on the calibrated frames, not
corrected for geometric distortions. We adopted the recommended parameters in 
the {\tt DOLPHOT} manual. The output magnitudes, in the VEGAMAG system, were 
corrected for aperture and  charge transfer efficiency (CTE).  The output position 
of each star is subsequently corrected for geometrical distortion using the reference frame.

In order to have a high quality CMD of \cl, the output parameters of 
{\tt DOLPHOT} were used to filter out non-stellar objects, cosmic rays and 
other spurious detections such as wings and diffraction rings. Basically 
the \textit{sharpness} parameter measures if the source is too sharp 
(positive values) or extended (negative values). The \textit{crowding} parameter 
measures how much the light of a star is due to close neighbours used in the 
PSF fit. As recommended in {\tt DOLPHOT}, no $\chi^{2}$ cut has been 
applied as a star selection criterion, due to its dependence on brightness. 
The following cuts were used: -0.1 $\leqslant$ sharpness 
$\leqslant$ 0.1; crowding $\leqslant$ 0.6; roundness $\leqslant$ 0.35. Only 
stars with type 1 and error type less than 8 in the {\tt DOLPHOT} 
classification system were kept. The total number 
of stars found with accurate photometry in both passbands is $\sim$ 32000. 

A sanity check was performed to verify the quality of our photometry. 
We applied the bandpass transformation given by \citet{sirianni} to 
convert the observed HB and main-sequence turn-off (MSTO)
magnitudes and colours from F606W and 
F814W to Johnson V and I. They were then compared to the CMD from
\citet{barbuy}. We have $V_{HB} = 16.40 \pm 0.03$ as the mean HB V band 
magnitude and associated error, whereas \citet{barbuy} got $V_{HB} = 16.35 \pm 0.04$. 
As for the MSTO, our CMD yields $(V,V-I)=(19.75,0.94)$, compared to $(V,V-I)=(19.7,0.95)$ 
from those authors. We thus conclude that both photometries agree
within the uncertainties.

\begin{figure}

\begin{center}
   \includegraphics[width=60mm,trim = 0 40 0 60]{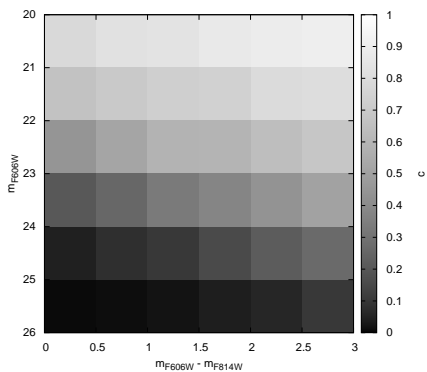}
   \caption{The position integrated completeness, $c(\mV, \mV - \mI)$ 
plotted over the CMD plane. The colour dependence is clear.}
   \label{comp}
\end{center}

\end{figure}

{\tt DOLPHOT} was employed again, this time in the artificial-star mode. 
Using fake-star experiments, we determined completeness as
a function of magnitude, colour and position in the cluster, 
$c(\mV,\mV - \mI,R)$. The task {\tt ACSFAKELIST} was used to generate 
artificial stars together with a Perl script. They were generated within the 
ranges $20 \leq \mV \leq 27$ mag and $0 \leq (\mV - \mI) \leq 2.5$. 
We adopted bin sizes of 1 mag in $\mV$, 0.5 mag in $(\mV - \mI)$ and 
100 pixels in distance from the cluster centre. The number of stars 
generated in a single realization of the experiment was never
larger than 10\% of the total original number in each region of the image. 
To build statistics, we performed 3 experiments for each $(\mV,\mV - \mI,R)$ 
bin. A total of $\sim~1.2 \times 10^{5}$ stars were generated. The
images with artificial stars added underwent the same
photometric and classification process as the original ones. As usual,
$c$ was taken as the ratio between the number of recovered artificial stars to
the total.
The result of the completeness analysis on the CMD plane is shown in 
Fig. \ref{comp}, where we sum the results over all the radii. 
There is a strong colour dependence in the expected sense:
at a fixed $\mV$, redder stars have higher $c$ values, since they are
brighter (and therefore more easily detected) in $\mI$.

\section{Analysis}

\subsection{Colour-Magnitude Diagram}

\begin{figure}
   \includegraphics[width=85mm]{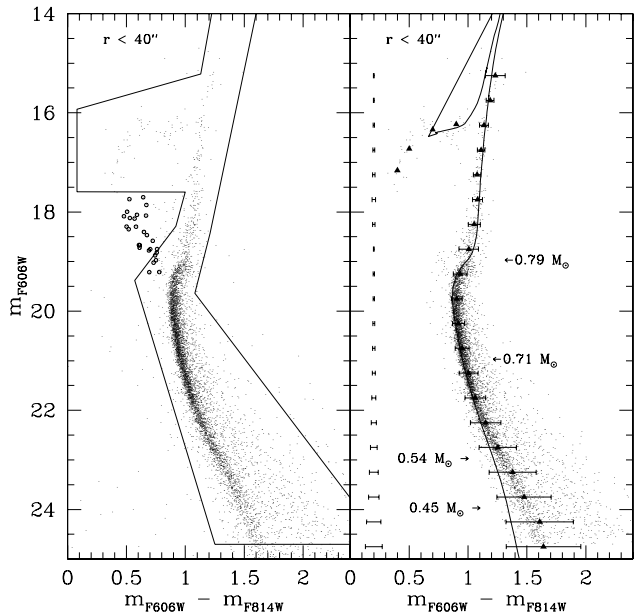}
   \caption{Left panel: CMD of all sources that satisfy the {\tt DOLPHOT} 
parameter cuts and which are located within $r \leqslant 40"$ from the
centre of \cl. We also show the region used to select the 
cluster evolutionary sequences. 
The BS candidates are marked with open circles. Right panel: same CMD as in the previous panel but cut according 
to the selected region. The solid triangles show the 
CMD fiducial line and corresponding dispersion. The mean photometric 
error is shown in the extreme left of this panel. Some mass values
are given along the MS. The best-fit 
isochrone is over plotted and has the following parameters: 
$log(age)=10.14$ ($\tau = 13.8$ Gyrs); $[Fe/H] = -1.80$; $E(B-V)= 0.46$; 
$\mM=14.55$; }
   \label{cmd}
\end{figure}

Fig. \ref{cmd} shows the CMD for the stars located within $r \leq 40"$
($\leq 1.6$ pc) from the nominal centre of \cl. By restricting the
image region, we strongly reduce field contamination.
The selected central region contains about 40\% of the total number 
of stars and only 10\% of the field area. 
Therefore the vast majority of the stars in this
field are expected to be GC members and should trace the evolutionary sequences 
of this object. In fact, we can see a well defined
structure in the main sequence (MS) spanning $\approx 6~mag$
below the MSTO (located at $\mV \simeq 19.2$). The sub-giant, red giant 
and horizontal branches 
(SGB, RGB and HB, respectively) are also clearly seen. Some candidate 
Asymptotic Giant Branch (AGB) stars are also present. The top of the 
RGB and AGB is not well sampled due to saturation, even
in the short exposures. 
The solid contour on the left panel indicates our selected limits 
for these stellar evolutionary loci. The chosen limits also 
avoid effects of saturation and incompleteness.
 For instance, our cut-off at $\mV \simeq 24.7$ at the
CMD bottom avoids stars with completeness corrections $1/c > 3$.
The open circles in the left panel indicate blue straggler
 (BS) candidates. 

On the right panel of Fig. \ref{cmd} the GC fiducial line is shown on
top of the selected stars. The mean colour was calculated in magnitude bins for 
the MS, SGB and RGB. In the HB, the fiducial sequence was 
built by computing the mean magnitude in colour bins. The sequence is 
shown as solid triangles. We also show the dispersion around the 
fiducial points plus the mean photometric error for each magnitude bin. 
The photometric errors decrease with brighter magnitudes as expected 
up to $\mV \simeq 19$. At yet brighter magnitudes, the uncertainties
initially increase because the measurements come from the short ACS exposures, 
since the long ones saturate at these bright levels. The dispersion around the 
fiducial points is slightly larger than the associated photometric errors. 
This is mostly due to the fact that the mean photometric errors are 
calculated over the entire image, while the SGB and RGB stars are mainly
located in the central region ($r \leqslant 40"$) of the cluster, where
crowding effects are not negligible. At the lower MS, unresolved binarism
play a role in causing additional spread in colours. The dispersion 
of the fiducial line is larger near the bright end of the MSTO due to field 
stars and BS contamination. The \cl inner field 
CMD is used to visually fit isochrones. 
We use the isochrone grid 
computed\footnote{http://stev.oapd.inaf.it/cgi-bin/cmd} by \citet{leo} 
for that purpose. We also allow for variations in $\ebv$ and $\mM$ in the
fits. The best-fit isochrone is shown on 
the right panel of Fig. 3. The corresponding parameters are 
$\lage = 10.14$ and $\feh  = -1.80$ , 
$\ebv = 0.46$ and $\mM = 14.55$.

The best-fit model describes most of the MS, including the 
MSTO, plus the SGB and RGB regions. 
A discrepancy is seen for $\mV > 23$, in the sense that the best-fit 
isochrone is too blue. This is
likely the result of unaccounted opacity in the low-mass 
Padova models \citep{baraffe}. The model HB is too short compared to the data, probably 
reflecting the model uncertainties at this evolutionary stage.
The best fit we find is a compromise between the two discrepancies just 
mentioned. For instance, in order to fit the cluster lower MS, a lower age 
must be used, but this eliminates the HB in the model, as we just
mentioned.
On the other hand, to better fit the HB, a very low metallicity must be 
adopted, which jeopardizes the fits to the other locci and also is in 
clear conflict with the spectroscopic metallicity determination by
\citet{minni}. 

In order to determine parameter uncertainties and to
test our visual fit, we generated model cluster fiducial lines and compared 
them with the observed one, shown in the right panel of Fig. \ref{cmd}. 
The models incorporated the effect of unresolved binaries, assuming a typical 
mass ratio of $m_2/m_1 = 0.5$ and binary fraction of 50\%. 
A $\chi^{2}$-like statistic was adopted in search of the best model and the 
parameter uncertainties were taken as the parameter range that satisfy the 
criterion $\chi^2_{min} \leqslant \chi^2 \leqslant 
2\times\chi^2_{min}$. This approach led to the following results: 
$\lage = 10.14 \pm 0.05$ and $\feh  = -1.80 \pm 0.2$ , $\ebv = 0.46 \pm 0.02$ 
and $\mM = 14.53 \pm 0.18$. We thus confirm the parameters of \cl using two 
different methods.

\subsection{Alternative age estimate}

In this section we explore an alternative way to constrain the age 
of \cl. A consistency check on the estimated age is 
obtained from the magnitude
difference between the HB and the MSTO, $\Delta V^{TO}_{HB}$. We here
adopt the same parametrization of this quantity as a function of 
$\feh$ and $\tau$ as used in Glatt et al (2008), 
which originally comes from Walker (1992):

\begin{equation}
log \tau(yrs) = -0.045~[Fe/H] + 0.37~\Delta V^{TO}_{HB} -0.24
\label{hbage}
\end{equation}

We estimate the HB and MSTO magnitudes as $V_{HB} \simeq 16.25$
and $V_{TO} \simeq 19.75$, respectively. Notice that the HB in special
has a considerable scatter around the quoted value. Assuming these 
values, we
obtain $\Delta V^{TO}_{HB} = 3.5$, yielding in turn an age $\tau = 13.7$
Gyrs. This estimate is consistent with that resulting from isochrone fitting.

\subsection{Density Profiles}

To analyse the density structure of \cl, we computed the 
Radial Density Profile (RDP). Stars at various radii are 
weighted by the inverse of their completeness and their
projected spatial density is obtained. The annuli used for the RDP 
construction were the same as in the completeness analysis. 

Building a meaningful RDP, however,
requires some cautionary steps. One is finding out the cluster centre,
which is determined by
the following method: (i) we count stars along the x-axis; (ii) the peak of
this distribution is then assumed to be the x coordinate of the
centre. The same procedure is then applied to the y-axis. For a spherical
distribution this procedure clearly yields the centre of the distribution of 
stars, which may also be assumed to be approximately the centre of mass. Adoption of alternative methods show that our centre determination is robust.

Another important issue is the effect of field boundaries. They cause 
areal completeness effects at the outer radial bins, which had to be taken 
into account in the density estimates. For that purpose,
we applied a Monte Carlo integral to find out the fraction of the
total area in each bin which was effectively sampled by the
image. Notice that in the 
proximity of the image boundary the error associated 
to the density estimate grows significantly due to the
increased loss of areal sampling. 

At each radius, the density of stars was given by:

\begin{equation}
\sigma(r) = \frac{\sum_{i} 1/c_{i}}{A(r)}
\label{denscomp}
\end{equation}

where the sum above takes place over all stars located within the
annulus of radius $r$, whose effective area (corrected for image
boundary) is $A(r)$. Each star is weighted by the
inverse of its associated completeness $c$ value, as estimated according to the
procedure described in Sect. 2.

\begin{figure}
  \includegraphics[width=80mm]{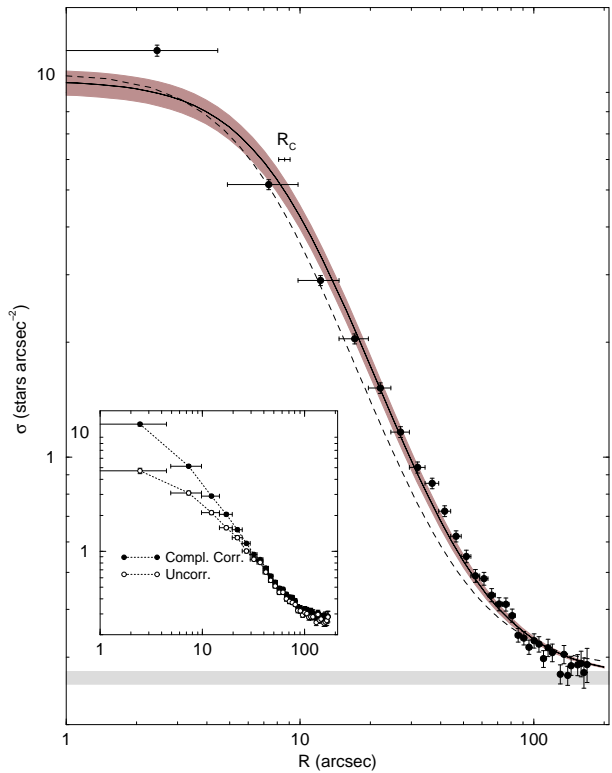}
  \caption{The RDP of \cl. We include only stars with $\mV \leqslant 24$. 
The filled circles show the completeness corrected number density
 profile, as given by eq. \ref{denscomp}. The horizontal bars show 
the radial range of each bin. A King-like profile fit is shown as the
solid line. The grey region is bracketed by the profiles
resulting from varying the best-fit parameters by $\pm 1 \sigma$. 
The dashed line is the best-fit profile assuming a higher value for the 
background star number density. The inset compares the completeness
corrected profile (solid points) with the uncorrected profile
(open symbols).
}
  \label{rdp}
\end{figure}

The resulting RDP is given as the solid circles in Fig. \ref{rdp}. The
horizontal bars correspond to the radial bin size. The solid
line is the best-fit King profile (see below). The $1 \sigma$ 
uncertainty range is also shown, as the grey region. 
In the inset we show the RDP that results from
counting all stars with the same weight (therefore not corrected for 
completeness). Clearly, completeness corrections are crucial, especially
in the central and denser regions, making the
RDP much steeper than the uncorrected one. The completeness
corrected RDP has no evidence for a core-like flattening.
A fit to a King-like profile yields 
$\rc = 8.6 \arcsec$ and $\sigma(0)=9.4\pm0.7$ stars $\arcsec^{-2}$.
The best-fit central density, however, is substantially lower than
the observed one. Changing the density of field stars in the fit does
not alter this conclusion.
We thus conclude that \cl is in fact a core-collapsed GC.

Using the inner field CMD shown in Fig. \ref{cmd} 
we can visually identify some BS. The number of candidate BS stars, as 
selected from their CMD 
position, is 50\% higher in the GC region than in the general field.
As the GC region represents only 10\% of the total image, the excess 
of BS candidates per unit area is a factor $\simeq 15$ higher than the 
field. This inequivocaly shows that these stars have a high probability 
of being cluster members. 

BS are known to be strongly concentrated towards the centre in most 
GCs, adding support to their possible origin from 
mergers of cluster stars or coalescence of tightly bound binaries. 
For the spatial study of the BS candidates we adopted the HB population
as a comparison set. By doing so we eliminate the dynamical
effects over the spatial distribution of stars since HB stars have a mass similiar
($m_{HB} \simeq 0.80 ~ M_{\odot}$) to the mass of the turn-off stars.

\begin{figure}
\begin{center}
  \includegraphics[width=65mm]{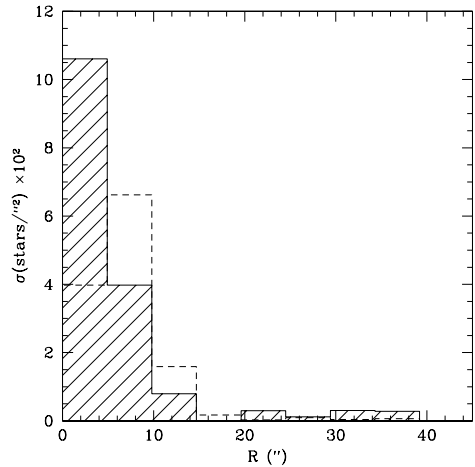}
  \caption{The density profile of selected BS (solid histogram) and HB 
(dashed histogram) stars, showing that most of the BS are located in the
 inner region of \cl. The BS population is more peaked than that of 
the HB stars.}
  \label{bs}
\end{center}
\end{figure}

In Fig. \ref{bs} we show the density
distribution of BS and HB stars as a function of distance from the GC centre. 
The BS stars are in fact strongly concentrated towards 
the inner region of NGC 6642. This highly concentrated BS distribution is the rule
in GCs, with the possible exceptions of NGC 2419 and $\omega$ Cen
\citep{dale1,ferraro}. These latter two, however, are 
atypical as GCs in many aspects, showing evidence for a complex star 
formation history and very large unrelaxed cores 
\citep{norris,sollima,bellazzini}. The distribution in
Fig. \ref{bs} also shows a region entirely depleted of BS 
candidates ($15 \arcsec < R < 20 \arcsec$) followed by
a residual population further out. This is consistent with what has been
observed in most GCs as discussed by \citet{dale2}.

\subsection{Luminosity and Mass Functions}

In this section we study the distribution of stars as a function of luminosity
and mass. These are important tools to assess dynamical effects, such as mass
segregation and stellar evaporation, that take place throughout the GC 
lifetime.

The luminosity function (LF) is computed in the same radial bins as in the
previous section. As in the case of RDPs, the LF is obtained by summing
over stars in a given annulus, but this time also as a function of
absolute magnitude $\MV$, as follows:

\begin{equation}
\phi(\MV) = \frac{dN}{d\MV} = \frac{\sum_i 1/c_i}{\Delta \MV}
\end{equation}

The sum above is carried out over all stars in the annulus whose
absolute magnitudes are within the bin around $\MV$ with width 
$\Delta \MV$. The $\MV$ values were computed by using the reddening
and $\mM$ derived in Sect. 3.1. Notice that we again weight stars according 
to the completeness function.

\begin{figure}
  \includegraphics[width=80mm]{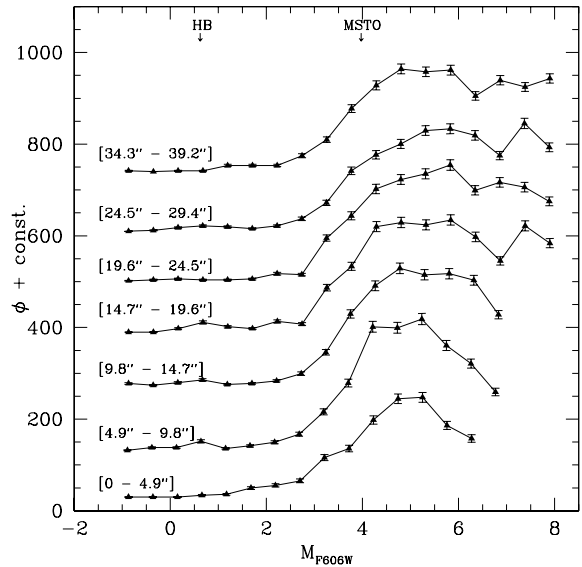}
  \caption{The LF of \cl stars at different radii, as indicated
next to each curve. 
The positions of the MSTO and HB are indicated on the top. An
offset was added to each curve to avoid overlap among them. In 
the inner regions, the completeness function falls off more rapidly as a 
function of magnitude, yielding a brighter cut-off limit.}
    \label{lf}
 \end{figure}

Fig. \ref{lf} shows the resulting LFs at different radial annuli.
$\Delta \MV = 0.5$ as adopted as bin size.
A constant offset was added to the curves to avoid cluttering.
Poisson error bars are also shown. The LFs extend towards 
fainter magnitudes at large radii because
the completeness levels are higher at these lower density regions.
Fig. \ref{lf} includes evolved
stars as well as MS ones. The most striking variation in the
LF shapes occurs in the MS domain ($\MV > 3.5$), where the
LF is clearly depleted of low luminosity stars and has a clear
peak in the inner cluster regions. In contrast, the outer regions
display flatter LFs all the way to the detection limit.

\begin{figure}
\begin{center}
  \includegraphics[width=80mm]{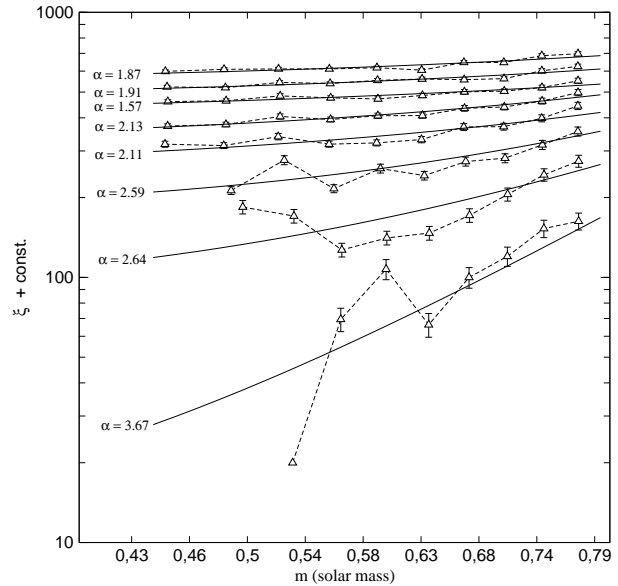}
  \caption{Present day mass functions (PDMFs) for different annuli 
around the centre of \cl. Distance from cluster centre increases upwards. 
We use the same radial bins used for the LFs. A constant offset was 
added to the star counts to avoid confusion. Fits to a power-law PDMF are
also shown and the corresponding slope values given. They are also listed
in Table \ref{lmctab}. }
  \label{mf}
\end{center}
\end{figure}

An alternative way to look for mass segregation is to compute the 
present day mass function (PDMF) from the observed LF. We use the 
best-fit Padova isochrone to convert stellar luminosities into initial 
masses. Only stars below $m \simeq 0.8 M_{\odot}$ are considered in this 
conversion, since 
according to the model, mass loss is more pronounced at higher masses, 
signalling evolutionary stages later than the MS. We then bin stars as a 
function of initial mass. The resulting
PDMFs are shown, again for the different radial annuli, in Fig. \ref{mf}.
A constant offset was again added for clarity. 
The star counts are normalized to unit solar mass intervals. Mass 
segregation now stands out more clearly, as the PDMF is systematically
steeper in the central regions of \cl. More striking, however, is the 
fact that the number of stars decreases towards lower masses, in contrast
to most observed PDMFs. Assuming a Initial Mass Function (IMF) with a
power-law behaviour, $\xi \propto m^{\alpha}$ ($\alpha = -2.35$ for a 
Salpeter IMF), the observed PDMF in \cl 
is evidence that this GC has lost most of its low mass stars to the field.
In order to quantify the slope of the observed PDMFs we carry out
power-law fits to the curves shown in Fig. \ref{mf}. The resulting
slopes $\alpha$ are shown in Table \ref{lmctab}. Notice again
the clear evidence for mass segregation, with a PDMF much more depleted
of low-mass stars in the central regions.

We should point out that this inverted PDMF slope has been 
observed in other recent studies. \citet{andre} found
a power-law slope $\alpha \simeq 0.9$ for NGC 6712 for masses $m \leq 0.8 
M_{\odot}$ and argue that this GC is very vulnerable to tidal disruption.
\citet{demar} also find a PDMF with decreasing number of
stars at lower masses ($\alpha \simeq 0.5$, in the range $0.2 \leq m/M_{\odot} 
\leq 0.8$) in NGC 2298. Finally, \citet{nath} obtain a
PDMF that peaks near the MSTO in NGC 6366.

Simulations of GC \citep{baumgardt} show that in tidal fields the 
preferential depletion of low-mass stars leads to a PDMF with inverted slope 
when the cluster has undergone 90\% or more of its associated dissolution time.
The time scale associated with the slope inversion phenomena is related
to the time when the compact stellar remnants start to dominate the cluster 
mass. Since \cl is a very old GC that resides in a central region of the Galaxy
the shape of the PDMF can easily be understood in the context of tidal interactions.

\begin{table}
\begin{center}
\begin{tabular}{lcc}

\hline
$\Delta r(\arcsec)$ & $m(M_{\odot})$ & $\alpha$\\
\hline
\hline
0 - 4.9     & 0.56 - 0.78 & $3.67 \pm 0.93$ \\
4.9 - 9.8   & 0.53 - 0.78 & $2.64 \pm 0.88$ \\
9.8 - 14.7  & 0.52 - 0.78 & $2.59 \pm 0.60$ \\
14.7 - 19.6 & 0.45 - 0.78 & $2.11 \pm 0.41$ \\
19.6 - 24.5 & 0.45 - 0.78 & $2.13 \pm 0.21$ \\
24.5 - 29.4 & 0.45 - 0.78 & $1.57 \pm 0.23$ \\ 
29.4 - 34.3 & 0.45 - 0.78 & $1.91 \pm 0.26$ \\
34.3 - 39.2 & 0.45 - 0.78 & $1.87 \pm 0.36$ \\
\hline
\end{tabular}
\caption{Results of the power-law fitting to the MF at 
different radii. Column 1 lists the radial bin in 
arcseconds; column 2 gives the selected mass limits for the 
analysis; column 3 shows fitted slopes $\alpha$.}
\label{lmctab}
\end{center}
\end{table}

\subsection{Horizontal Branch Morphology}

We here determine the HB morphology using the parametrization described 
in \citet{lee} 

\begin{equation}
\hbi = \frac{B-R}{(B + R + V)}
\end{equation}

where B is the number of stars in the blue part of the HB, R 
is the number of stars in the red part and V is the number of variable 
stars. To determine the colour of the RR-Lyrae instability strip we used 
the HB analysis of the halo GC NGC 4147 by \citet{stet2}.
This cluster has a $\feh$ very similar to \cl. \citet{stet2} lists
mean BVI magnitudes and colours for a sample of well known RR-Lyrae 
in NGC 4147. We used the full range of mean colours as the instability
strip. Our photometry was converted to the standard system by means of
the transformations from \citet{sirianni}. The resulting index is then
$\hbi \simeq 0.25$. 

An alternative approach to determine $\hbi$ is to use the 
colour range of the instability strip quoted by \citet{mac}, based on
the work of \citet{smith} and on the data from \citet{pio}.
These limits are in $(B-V)$ and were converted to $(V-I)$ using 
colour-colour tranformations from \citet{cald}. The result
is then $\hbi = 0.23$. This is in close agreement with the previous
estimate. We use the uncertainties of $0.02$ mag in the stability strip 
limits quoted by \citet{mac} and propagate them into $\hbi$. Our final
result is then $\hbi = 0.23^{+0.05}_{-0.01}$.
This value is in apparent disagreement with that calculated by 
\citet{mac} for \cl using the same limits but applied to the published CMDs 
from \citet{pio}: $\hbi = -0.04 \pm 0.14$. This discrepancy, however,
is dominated by the different reddening corrections adopted.
We use $\ebv = 0.46$ whereas \citet{pio} find $\ebv = 0.41$. 
A lower E(B-V) systematically shifts the entire HB towards redder
intrinsic colours. This certainly yields a lower index. If we
adopt this smaller E(B-V) value to correct our observed HB and
again apply the same colour range as \citet{mac}, we obtain
$\hbi = 0.05^{+0.07}_{-0.09}$.

\begin{figure}
\begin{center}
  \includegraphics[width=80mm]{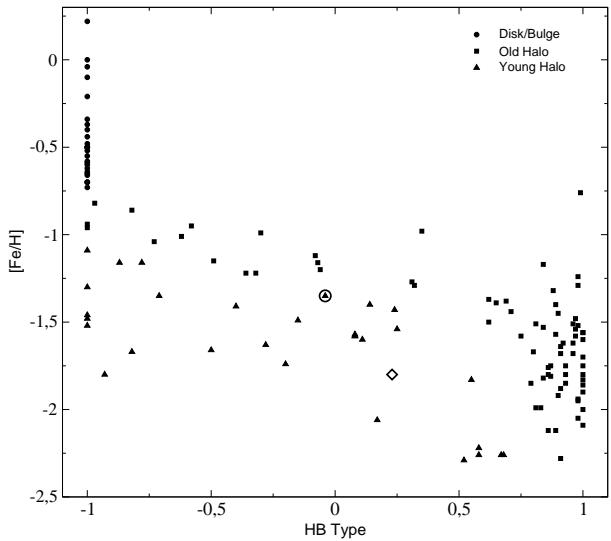}
  \caption{$\hbi$ vs. $\feh$ diagram from \citet{mac}. Here we show 
the Old Halo clusters(solid boxes), Young Halo clusters(solid triangles) 
and the Disk/Bulge clusters(solid circles). The solid triangle inside the open 
circle indicates the original position of \cl while the open diamond shows 
the position found in this work.}
  \label{hbfeh}
\end{center}
\end{figure}

With $\hbi$ at hand, we may plot \cl in the HB-type vs. metallicity
diagram. \citet{mac} use $\feh = -1.35$ for \cl and place it as
a member of the Young Halo subsystem, according to the classification of 
\citet{bergh}. Our $\hbi$ and $\feh$ estimates 
confirm this position as far as the HB morphology is concerned,
displacing \cl even further from the Old Halo and Disk/Bulge loci as seen in Fig. \ref{hbfeh}.
On the other hand, its age and position in the Galaxy suggest that \cl 
is more likely an inner halo than an outer bulge transition object. 

\section{Discussion}

We studied the GC \cl with ACS/HST imaging in the F606W and F814W bands
using combined long and short exposures. The ACS/HST photometry employed 
here makes it possible to resolve stars well in the vicinity of the 
cluster's collapsed core. 

From the GC CMD we have derived an age of $13.8 \pm 1.6 ~ Gyrs$ 
and $\feh = -1.80 \pm ~ 0.20$. We also derived a colour excess 
of $\ebv = 0.46 \pm 0.02$ towards \cl and a distance from
the Sun comparable to that of the Galactic centre, $\ds = 8.05 \pm 0.66 ~ 
kpc$ object. Given the direction of \cl on the sky, and assuming
that the distance from the Sun to the Galactic Centre is $d_{C,\odot} = 8.0$
kpc, we conclude that \cl in fact currently occupies a very central 
position in the Galaxy, $D_{C,clus} = 1.4$ kpc.

\cl has a very dense central core. Therefore,
it is likely to be composed of stars that have undergone 
(or are undergoing) mergers. The very dense central object seen in 
Fig. \ref{cluster} in not well resolved, although a high fraction of the BS
are found in its vicinity. The radial density profile 
shows no evidence of a well resolved core, corroborating the idea
that \cl is a core-collapsed GC. 

A number of other surrounding BS
candidates has been found. Their spatial distribution 
is more concentrated towards the GC centre than other GC stars, such as 
those in the HB. This result can also be interpreted in the context 
of the BS formation by merger scenario. We also observe a hint of bimodal 
radial density profile of the BS candidates, with a depletion region
in the middle. These are common features in GCs \citep{dale2}. 

The studied CMD also exhibits a well developed HB from which
we could estimate $\hbi$ as defined in \citet{lee}. Along with the
 $\lage$ and $\feh$ values found in this work, our analysis
shows that \cl is an unusual GC. It has an old age and a central location
in the Galaxy, but its position in the $\feh$ and $\hbi$ plane is
more consistent with the young halo GC system.
Having in mind its atypical behaviour concerning its dynamical evolution 
and HB morphology, \cl is more likely to be a of a transition class 
between the inner halo and outer bulge cluster population.

The analysis of the LF and PDMF and their variation as a function 
of radius shows a clear evidence for mass segregation,
especially in the range $0.4 \leq m/m_{\odot} \leq 0.8$. Furthermore,
in all regions, there is a decrease in the star counts towards lower
luminosities and masses, an effect which is stronger in the central
regions. This inversion in the PDMF slope
is atypical, although it has been previously observed and 
is supported by dynamical N-body simulations \citep{andre,demar,nath}. 
Our findings concerning
the PDMF of \cl, which resides in a violent environment, can be 
explained in terms of disk and bulge shocking through its perigalacticon 
passages \citep{baumgardt}. The high resolution images available 
nowadays are allowing us to observe these phenomena more frequently and 
strongly suggest that dynamical effects over a GC
life-time have a dominant contribution from the external environment. 

An interesting feature that is expected to result from the
high degree of depletion of low mass stars is the presence
of a tidal tail around \cl. For this purpose, a more sophisticated 
field decontamination technique must be adopted such as the statistical 
analysis of CMDs \citep{kerber}. From the analysis of the RDP of \cl 
it is not clear that we have reached the tidal radius of the
GC; thus, for the analysis of a possible tidal tail it is necessary to 
have images with a wider field. The tidal tail search could make the 
estimate of the mass loss due to tidal erosion possible.

{\bf Acknowlegments.} This work was supported by Conselho Nacional
de Desenvolvimento Cient\'\i fico e Tecnol\'ogico (CNPq) in Brazil.

\label{lastpage}
\end{document}